\documentstyle[preprint,aps,epsfig]{revtex}
\draft
\begin{document}
\preprint{Phys. Rev. Lett. {\bf 88} (2002) 192701}
\title{Probing the High Density Behaviour of Nuclear Symmetry Energy
 with High Energy Heavy-Ion Collisions}
\bigskip
\author{\bf Bao-An Li\footnote{email: Bali@astate.edu}}
\address{Department of Chemistry and Physics\\
P.O. Box 419, Arkansas State University\\
State University, Arkansas 72467-0419, USA}
\maketitle

\begin{quote}
High energy heavy-ion collisions are proposed as a novel means to 
constrain stringently the high density (HD) behaviour of 
nuclear symmetry energy. Within an isospin-dependent hadronic transport 
model, it is shown for the first time that the isospin asymmetry of 
the HD nuclear matter formed in high energy heavy-ion collisions is uniquely 
determined by the HD behaviour of nuclear symmetry energy. 
Experimental signatures in two sensitive probes, i.e., $\pi^-$ to $\pi^+$ ratio 
and neutron-proton differential collective flow, are also investigated.\\
{\bf PACS} numbers: 25.70.-z, 25.75.Ld., 24.10.Lx
\end{quote}

\newpage
The high density (HD) behaviour of nuclear symmetry energy $E_{sym}(\rho)$ 
is very important for understanding many interesting astrophysical phenomena, 
but it also subjects to the worst uncertainty among all properties of dense 
nuclear matter\cite{kut}. To study the HD behaviour of $E_{sym}(\rho)$ 
has been a longstanding goal of extensive research with various 
microscopic and/or phenomenological models over the last few decades, e.g., 
\cite{hdsym}, for a recent review, see, e.g. \cite{bom01}. The model 
predictions are extremely diverse and often contradictory. The results can 
be roughly classified into two groups, i.e., a group where the $E_{sym}(\rho)$ 
rises and one in which it falls with 
the increasing density. For instance, within the Hartree-Fock approach 
using about 25 Skyrme and Gony effective interactions that have been widely 
used successfully in studying saturation properties of symmetric nuclear 
matter and nuclear structures near the $\beta$ stability valley, the calculated 
symmetry energies were found to fall approximately equally into the 
two groups\cite{brown,mar}. The density dependence of nuclear symmetry energy, 
especially at high densities, has many profound consequences for 
various studies in astrophysics\cite{lat00,pra97}. In particular, 
an increasing $E_{sym}(\rho)$ leads to a relatively more 
proton-rich neutron star whereas a decreasing one would make the neutron star 
a pure neutron matter at high densities. Consequently, the chemical composition 
and cooling mechanisms of protoneutron stars\cite{lat91,sum94}, critical densities for
Kaon condensations in dense stellar matter\cite{lee96,kurt2}, mass-radius 
correlations\cite{prak88,eng94} as well as the possibility of a mixed 
quark-hadron phase\cite{kurt3} in neutron stars will all be rather different. 
The fundamental cause of the extremely uncertain HD behaviour of 
$E_{sym}(\rho)$ is the complete lack of terrestrial laboratory 
data to constrain directly the model predictions. In this Letter, 
high energy heavy-ion collisions are proposed as a novel means to constrain 
stringently the HD behaviour of $E_{sym}(\rho)$. The upper bound of HD symmetry 
energy effects on high energy heavy-ion collisions is explored.
It is shown for the first time that the neutron/proton ratio of HD nuclear 
matter formed in high energy heavy-ion collisions is uniquely determined 
by the HD behaviour of $E_{sym}(\rho)$. Experimental signatures of
the HD behaviour of $E_{sym}(\rho)$ in two sensitive observables, 
i.e., the $\pi^-/\pi^+$ ratio and the neutron-proton differential 
collective flow, are also investigated. 

We use here the parabolic approximation of the
equation of state (${\rm EOS}$) for isospin asymmetric nuclear matter\cite{bom01} 
\begin{equation}\label{ieos}
e(\rho,\delta)= e(\rho,0)+E_{sym}(\rho)\delta^2,
\end{equation}
where $e(\rho,0)$ is the energy per nucleon in symmetric nuclear matter, 
and $\delta\equiv (\rho_n-\rho_p)/(\rho_n+\rho_p)$ is the isospin asymmetry
in terms of the neutron ($\rho_n$) and proton ($\rho_p$) density, respectively.
For the isoscalar part we use the simplest, momentum-independent parameterization
\begin{equation}
e(\rho,0)= \frac{a}{2}u+\frac{b}{1+\sigma}u^{\sigma}+\frac{3}{5}e_F^0u^{2/3},
\end{equation}
where $u\equiv \rho/\rho_0$ is the reduced density and $e_F^0=36$ MeV is the Fermi 
energy. The parameters $a=-358.1$ MeV, $b=304.8$ MeV and $\sigma=7/6$ 
are determined by saturation properties and a compressibility 
$K_{\infty}=201$ MeV of isospin symmetric nuclear matter. 
Our conclusions in this work are independent of the particular 
form of $e(\rho,0)$. The symmetry energy is\cite{kut,bom01}
\begin{equation}
E_{sym}(\rho)\equiv e(\rho,1)-e(\rho,0)=\frac{5}{9}E_{kin}(\rho,0)+V_2(\rho), 
\end{equation}
where $E_{kin}(\rho,0)$ is the kinetic energy per nucleon in symmetric nuclear 
matter and $V_2(\rho)$ is the deviation of the interaction energy of 
pure neutron matter from that of symmetric nuclear matter. 
The $E_{sym}(\rho)$ becomes negative if the condition 
$V_2(\rho)\leq -\frac{5}{9}E_{kin}(\rho,0)$ is reached at high densities. 
Thus a pure neutron matter could become most stable, leading to the 
isospin separation instability in HD neutron-rich matter. 
Consequently, pure neutron domains or neutron 
bubbles surrounding isolated protons can be formed in 
neutron stars\cite{kut}. To represent the two groups
of model predictions, we use for the $E_{sym}(\rho)$ 
\begin{equation}\label{esym}
E^a_{sym}(\rho)\equiv E_{sym}(\rho_0)u
{\rm ~and~} 
E^b_{sym}(\rho)\equiv E_{sym}(\rho_0)u\cdot\frac{u_c-u}{u_c-1},
\end{equation}
where $E_{sym}(\rho_0)=30$ MeV is the symmetry energy at normal nuclear matter
density, and $u_c=\rho_c/\rho_0$ is the reduced critical density at 
which the $E^b_{sym}(\rho)$ crosses zero and becomes negative at 
higher densities. The predicted value of $u_c$ ranges from 
about 2.7 (Hartree-Fock with the Skyrme interaction Sp\cite{mar}) 
to 9 (variational many-body approach with the
UV14+UVII interaction\cite{hdsym}.). These two forms of the
$E_{sym}(\rho)$ (insert in the lower window) together with the corresponding 
${\rm EOS}$ for isospin asymmetric nuclear matter are shown in Fig.\ 1. 
With the linearly increasing $E^a_{sym}(\rho)$, the ${\rm EOS}$ (upper window) 
becomes stiffer with the increasing $\delta$. The isospin symmetric nuclear matter 
remains to be the ground state at all densities. This is in stark contrast to the
situation using the $E^b_{sym}(\rho)$. The ${\rm EOS}$ obtained with 
the $E^b_{sym}(\rho)$ and $u_c=3$ (lower window), is softened instead of being
stiffened by the increasing isospin asymmetry $\delta$ at densities 
higher than $3\rho_0$. At these high densities, the pure neutron matter 
and the isospin symmetric nuclear matter thus becomes the most 
stable and unstable form of HD nuclear matter, respectively.

High energy heavy-ion collisions provide the only terrestrial situation 
where the HD neutron-rich matter can be formed. Moreover, fast 
radioactive heavy-ion beams to be available at the planned Rare Isotope 
Accelerator (RIA) in the United States can make the HD nuclear 
matter even more neutron-rich. The HD behaviour of $E_{sym}(\rho)$ 
affects properties of the HD nuclear matter formed in high energy 
heavy-ion collisions. Moreover, it leads to interesting precursory 
phenomena in the experimental observables of high energy heavy-ion collisions. 
We investigate these effects and phenomena within an 
isospin-dependent hadronic transport model\cite{ibuu1,ibuu2}. 
Evolutions of the phase-space distribution functions of nucleons, 
Delta resonances and pions with their explicit isospin degrees 
of freedom are solved numerically by using of the test-particle 
approach\cite{wong,bet}. Both the isoscalar and isovector 
mean-filed potentials are derived consistently from 
the nuclear ${\rm EOS}$ given above. Isospin-dependent 
total and differential cross sections among all particles are 
taken either from the elementary particle scattering data 
or obtained by using the detailed balance. Explicitly 
isospin-dependent Pauli blockings for fermions are also employed. For a 
review of the model, we refer the reader to ref.\cite{lkb98}. 

Shown in Fig.\ 2 is the correlation between the isospin asymmetry $\delta$ and 
the baryon density $\rho/\rho_0$ at the instant of 20 fm/c in the reaction 
of $^{132}Sn+^{124}Sn$ at a beam energy of 400 MeV/nucleon and an impact 
parameter of 1 fm. This very reaction will be available at the RIA facility. 
The overall rise of $\delta$ at very low densities is due to the
neutron skins of the colliding nuclei. One also notices that 
the $E^{b}_{sym}(\rho)$ leads to a slightly higher compression 
due to its relatively softening effect on the nuclear ${\rm EOS}$. 
Dramatic effects due to the different symmetry energies 
are clearly revealed especially at high densities. For a comparison, 
the corresponding correlation in neutron stars at $\beta$ equilibrium 
is shown in the insert. As a reference, an estimate of the direct URCA 
limit of $\delta_{\beta}$ (below which the fast cooling process 
can happen in protoneutron stars and thus results in also higher neutrino flux) 
is indicated with the dotted line\cite{lat91}. To a good approximation 
perfectly suitable for this study, a neutron star can be considered 
as consisting of nuetrons, protons and electrons. At $\beta$ equilibrium 
the proton fraction $x_{\beta}\equiv \rho_p/(\rho_n+\rho_p)$ in 
neutron stars is then determined by\cite{lat91}    
\begin{equation}\label{fraction}
\hbar c(3\pi^2\rho x_{\beta})^{1/3}=4E_{\rm sym}(\rho)(1-2x_{\beta}).
\end{equation}   
The equilibrium isospin asymmetry $\delta_{\beta}=1-2x_{\beta}$ is therefore 
entirely determined by the $E_{sym}(\rho)$. With the $E^{b}_{sym}(\rho)$, 
the $\delta_{\beta}$ is $1$ for $\rho/\rho_0\geq 3$, 
indicating that the neutron star has become a pure neutron matter at 
these high densities. On the contrary, with the $E^{a}_{sym}(\rho)$, 
the neutron star becomes so proton-rich that the direct 
URCA process can happen at densities higher than about $2.3\rho_0$. 
An astonishing similarity is seen in the resultant 
$\delta-\rho$ correlations for the neutron star and the heavy-ion collision. 
In both cases, the symmetry energy $E^b_{sym}(\rho)$ makes the HD nuclear 
matter more neutron-rich than the $E^a_{sym}(\rho)$ and the effect 
grows with the increasing density. Of course, this is no surprise 
since the same nuclear ${\rm EOS}$ is at work in both cases. 
As shown in Fig.\ 1, in stark contrast to the situation with 
the $E^a_{sym}(\rho)$, the decreasing $E^b_{sym}(\rho)$ above $\rho_0$ 
makes it energetically more favorable to have the denser region more 
neutron-rich. 

To further investigate the HD behaviour of $E_{sym}(\rho)$, 
the average $n/p$ ratio of the HD region with $\rho/\rho_0\geq 1$ 
is examined as a function of time and beam energy in 
the left panels of Fig.\ 3. The effect on $(n/p)_{\rho/\rho_0\geq 1}$ 
due to the different $E_{sym}(\rho)$ is seen to grow with both the reaction
time and the beam energy. This is because of the higher densities
reached with the more energetic beams and after longer times of compression. 

How to probe experimentally the HD behaviour of $E_{sym}(\rho)$ 
in high energy heavy-ion collisions? Among the observables 
we have explored, the $\pi^-/\pi^+$ ratio and the neutron-proton 
differential collective flow are found to be most promising. The former 
measures sensitively the neutron to proton ratio $n/p$ of the 
HD nuclear matter and thus indirectly the HD behaviour of 
$E_{sym}(\rho)$. At beam energies below about 2 GeV/nucleon, 
pions are mostly produced through the decay of $\Delta(1232)$ 
resonances. The primordial $\pi^-/\pi^+$ ratio is approximately quadratic 
in n/p according to the branching ratios of single pion 
production via $\Delta$ resonances in nucleon-nucleon collisions
\begin{equation}
\pi^-/\pi^+=\frac{5n^2+np}{5p^2+np}\approx (n/p)^2.
\end{equation}
Pion reabsorptions and rescatterings are expected to 
complicate the above relationship. Nevertheless, 
very high sensitivity to the $n/p$ ratio is retained in 
the final $\pi^-/\pi^+$ ratio as indicated in the experimental 
data of high energy heavy-ion collisions\cite{stock}. 
Shown in the right panels of Fig.\ 3 are the $(\pi^-/\pi^+)_{like}$ 
ratio 
\begin{equation}
(\pi^-/\pi^+)_{like}\equiv \frac{\pi^-+\Delta^-+\frac{1}{3}\Delta^0}
{\pi^++\Delta^{++}+\frac{1}{3}\Delta^+}
\end{equation} 
as a function of time. This ratio naturally becomes the final 
$\pi^-/\pi^+$ ratio when the reaction time $t$ is much longer 
than the lifetime of the delta resonance $\tau_{\Delta}$.
The $(\pi^-/\pi^+)_{like}$ ratio is rather high in the early 
stage of the reaction because of the large numbers of neutron-neutron 
scatterings near the surfaces of the colliding nuclei. 
By comparing the two sides in Fig.\ 3, 
it is seen that a variation of about 30\% in the 
$(n/p)_{\rho/\rho_0\geq 1}$ ratio due to the different $E_{sym}(\rho)$ 
results in about 15\% change in the final $\pi^-/\pi^+$ ratio.
The later thus has an appreciable response factor of 
about 0.5 to the variation of HD n/p ratio, and it is approximately 
beam energy independent. 

The neutron-proton differential collective flow is measured
by\cite{li00} 
\begin{equation}
F_{np}(y)\equiv\frac{1}{N(y)}\sum_{i=1}^{N(y)}p_{x_i}\tau_i,
\end{equation}
where $N(y)$ is the total number of free nucleons at the rapidity $y$, 
$p_{x_i}$ is the transverse momentum of particle $i$ in the reaction 
plane, and $\tau_i$ is $+1$ and $-1$ for neutrons and protons, respectively.
The $F_{np}(y)$ combines constructively the in-plane transverse 
momenta generated by the isovector potentials while reducing 
significantly influences of the isoscalar potentials of both neutrons 
and protons. Thus, it can reveal more directly the HD behaviour 
of $E_{sym}(\rho)$ in high energy heavy-ion collisions. A typical 
result for the $^{132}Sn+^{124}Sn$ reaction is shown in the upper 
window of Fig.\ 4. A clear signature of the HD behaviour of $E_{sym}(\rho)$ 
appears at both forward and backward rapidities. To characterize 
the effect, the slope $dF_{np}/d(y_{cm}/y_{beam})$ 
at mid-rapidity is shown as a function of beam energy in the lower 
window. A striking difference of about a factor of 2 exists in the 
reactions at $E_{beam}\geq 200$ MeV/nucleon. This large effect 
can be very easily observed by using available detectors at several heavy-ion 
facilities in the world\cite{lyn,gary}. Compared with the $\pi^-/\pi^+$ ratio, 
the neutron-proton differential collective flow is more directly affected 
by and is thus also a more sensitive probe of the HD behaviour 
of $E_{sym}(\rho)$. 

In conclusion, the HD behaviour of $E_{sym}(\rho)$ has been 
puzzling physicists for decades. In this work, high energy 
heavy-ion collisions are proposed as a novel means 
to solve this longstanding problem. The upper bound of HD symmetry 
energy effects on high energy heavy-ion collisions is explored.
For the first time, it is shown that the isospin asymmetry of HD 
nuclear matter formed in high energy heavy-ion collisions is 
uniquely determined by the HD behaviour of $E_{sym}(\rho)$. 
Both the $\pi^-/\pi^+$ ratio and the neutron-proton differential 
collective flow can serve as sensitive probes. Measurements of these 
observables will provide the first terrestrial data to constrain stringently 
the HD behaviour of nuclear symmetry energy. 

This work was supported in part by the National Science Foundation Grant 
No. PHY-0088934 and Arkansas Science and Technology Authority Grant No. 00-B-14.

\newpage
\begin{figure}[htp] 
\vspace{3.5cm}
\centering \epsfig{file=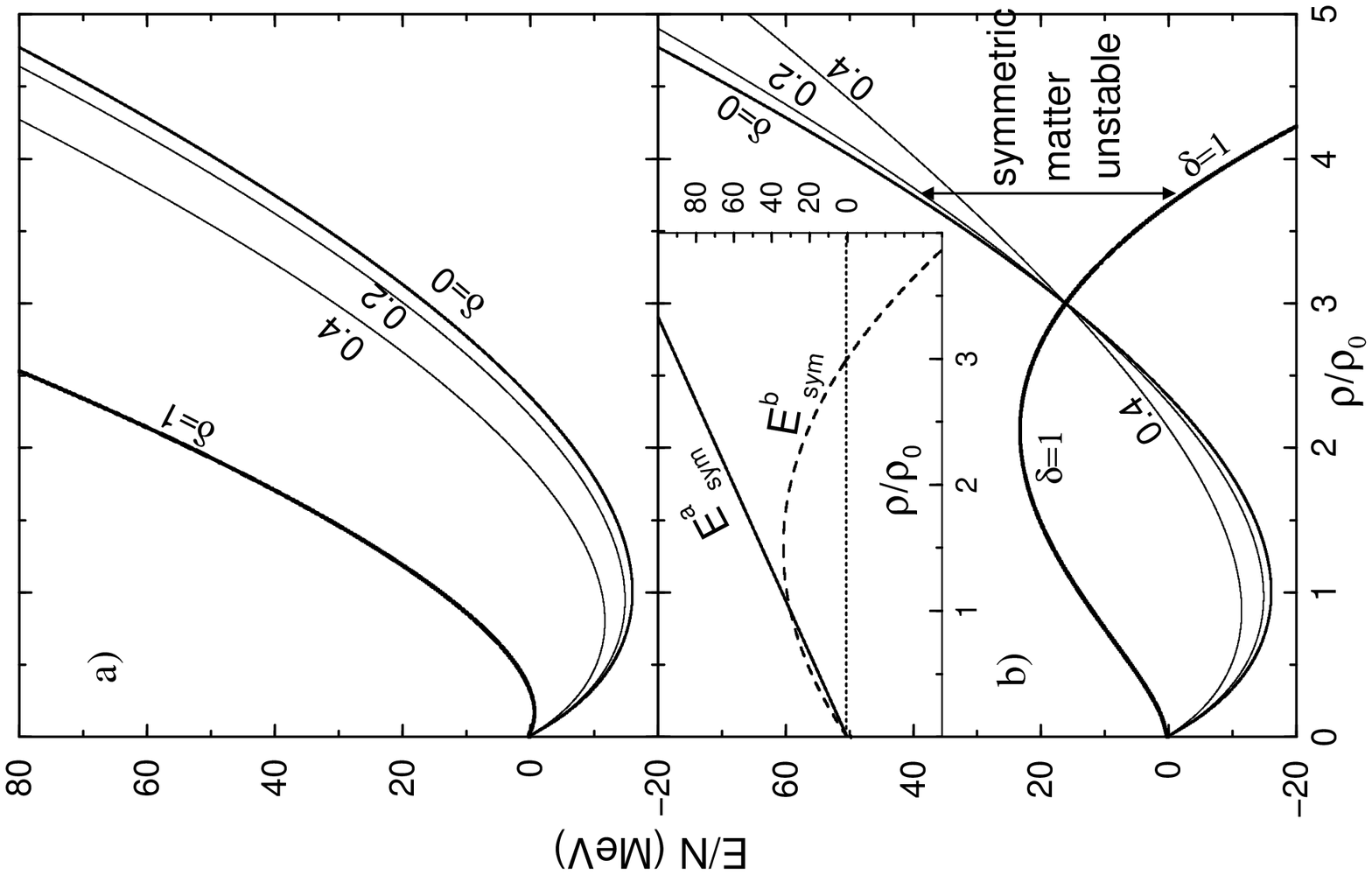,width=14cm,height=10cm,angle=-90} 
\vspace{1.cm}
\caption{The upper (lower) window is the equation of state of 
isospin-asymmetric nuclear matter corresponding to the 
nuclear symmetry energy parameterization $E^a_{sym}$ ($E^b_{sym})$ 
shown in the insert of the lower window.} 
\label{fig1}
\end{figure}

\begin{figure}[htp] 
\vspace{3.5cm}
\centering \epsfig{file=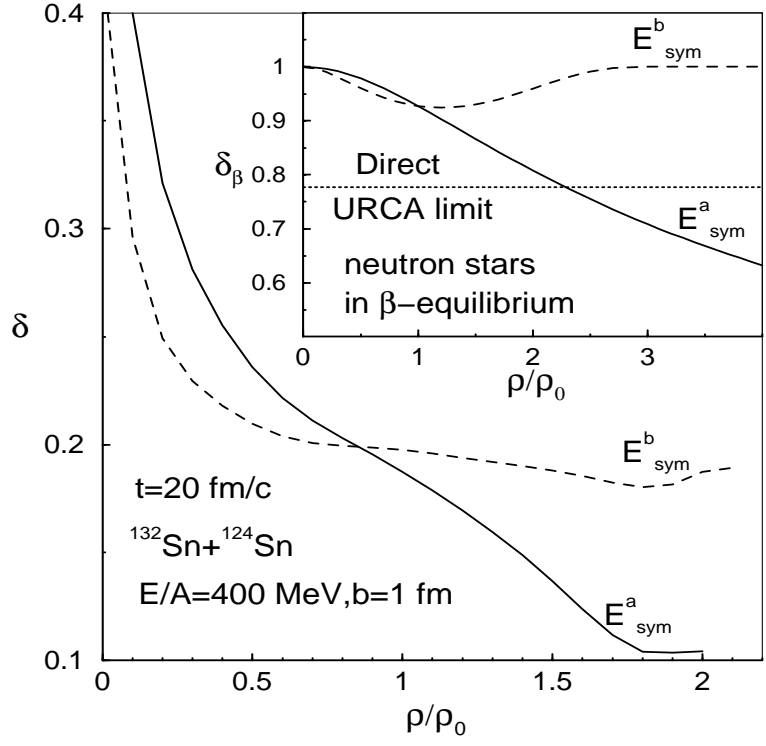,width=10cm,height=10cm,angle=-90} 
\vspace{1.cm}
\caption{The isospin asymmetry-density correlation at t=20 fm/c over the
whole space in the $^{132}Sn+^{124}Sn$ reaction with the nuclear symmetry 
energy $E^a_{sym}$ and $E^b_{sym}$, respectively. The corresponding correlation 
in neutron stars is shown in the insert.} 
\label{fig2}
\end{figure}

\begin{figure}[htp]
\vspace{3.5cm} 
\centering \epsfig{file=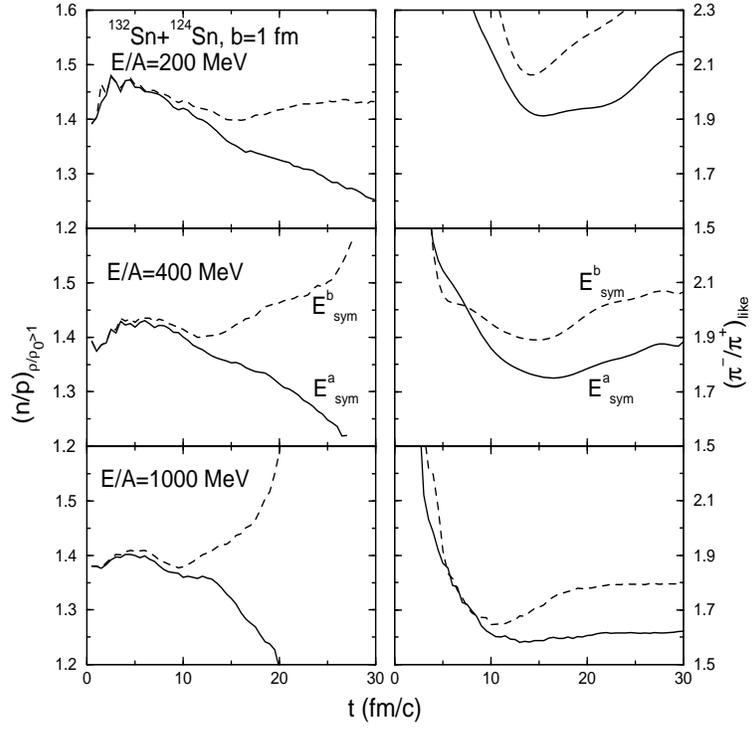,width=10cm,height=10cm,angle=-90} 
\vspace{1.cm}
\caption{Left panels: the neutron/proton ratio of nuclear matter with density
higher than normal nuclear matter density as a function of time
with the nuclear symmetry energy $E^a_{sym}$ and $E^b_{sym}$, respectively. 
Right panels: $(\pi^-/\pi^+)_{like}$ ratios in the same reactions.} 
\label{fig3}
\end{figure}

\begin{figure}[htp] 
\vspace{3.5cm}
\centering \epsfig{file=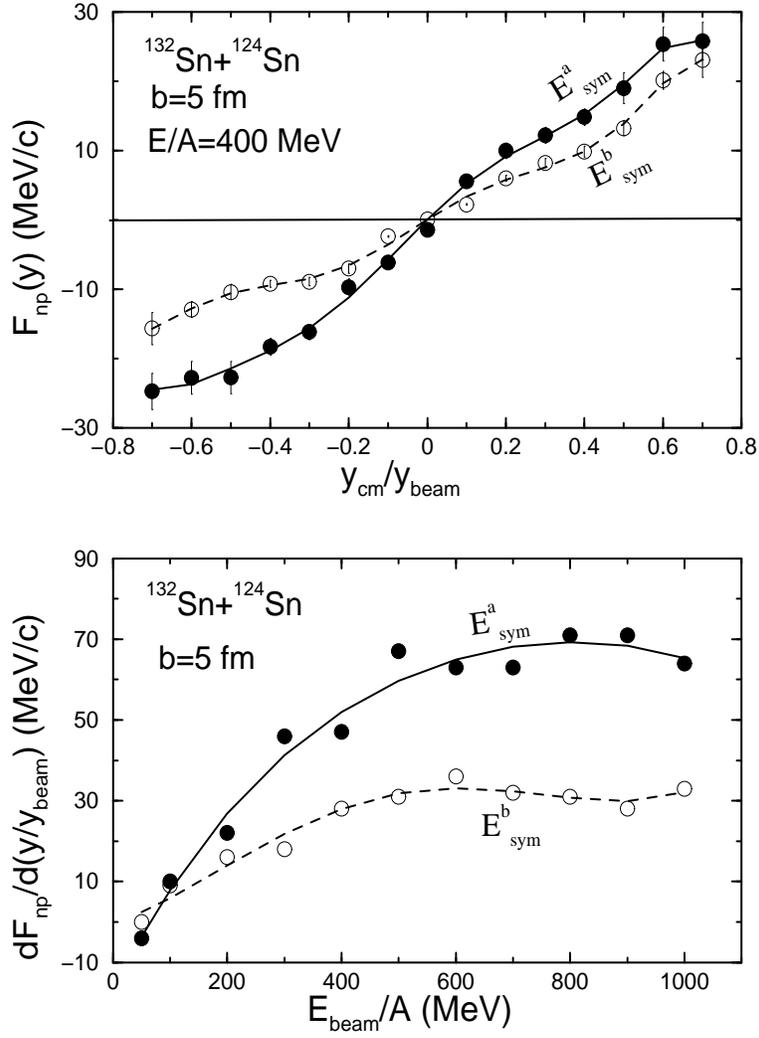,width=14cm,height=10cm,angle=-90} 
\vspace{1.cm}
\caption{Upper window: the neutron-proton differential collective flow
in the $^{132}Sn+^{124}Sn$ reaction with the nuclear symmetry energy 
$E^a_{sym}$ and $E^b_{sym}$, respectively. Lower window: excitation function
of the slope parameter of the differential flow for the $^{132}Sn+^{124}Sn$ reaction.
The solid (with the $E^a_{sym}$) and dashed (with the $E^b_{sym}$) lines are drawn 
to guide the eye.}
\label{fig4}
\end{figure}

\end{document}